\documentclass[reprint,superscriptaddress,amsmath,amssymb,prb,]{revtex4-1}
\usepackage{graphicx}
\usepackage{epstopdf}
\usepackage{dcolumn}
\usepackage{bm}
\usepackage{epstopdf}
\begin{document}
\title{Probing Magnetic Sublattices in Multiferroic Ho$_{0.5}$Nd$_{0.5}$Fe$_{3}$(BO$_{3}$)$_{4}$ Single Crystal\\ using X-ray Magnetic Circular Dichroism}

\author{Mikhail Platunov}
\email{ms-platunov@yandex.ru}
\affiliation{Kirensky Institute of Physics, Federal Research Center KSC SB RAS, Akademgorodok 50, bld. 38, 660036 Krasnoyarsk, Russia}
\affiliation{ESRF-The European Synchrotron, 71 Avenue des Martyrs CS40220, F-38043 Grenoble Cedex 9, France.}

\author{Natalia Kazak}%
\affiliation{Kirensky Institute of Physics, Federal Research Center KSC SB RAS, Akademgorodok 50, bld. 38, 660036 Krasnoyarsk, Russia}

\author{Viacheslav Dudnikov}%
\affiliation{Kirensky Institute of Physics, Federal Research Center KSC SB RAS, Akademgorodok 50, bld. 38, 660036 Krasnoyarsk, Russia}

\author{Fabrice Wilhelm}%
\author{Amir Hen}%
\author{Vadim Diadkin}%
\author{Iurii Dovgaliuk}%
\author{Alexey Bosak}%
\affiliation{ESRF-The European Synchrotron, 71 Avenue des Martyrs CS40220, F-38043 Grenoble Cedex 9, France.}

\author{Vladislav Temerov}%
\author{Irina Gudim}%
\author{Yurii Knyazev}%
\affiliation{Kirensky Institute of Physics, Federal Research Center KSC SB RAS, Akademgorodok 50, bld. 38, 660036 Krasnoyarsk, Russia}

\author{Sergey Gavrilkin}%
\affiliation{P.N. Lebedev Physical Institute of RAS, 119991 Moscow, Russia}

\author{Andrei Rogalev}%
\affiliation{ESRF-The European Synchrotron, 71 Avenue des Martyrs CS40220, F-38043 Grenoble Cedex 9, France.}

\author{Sergei Ovchinnikov}%
\affiliation{Kirensky Institute of Physics, Federal Research Center KSC SB RAS, Akademgorodok 50, bld. 38, 660036 Krasnoyarsk, Russia}
\begin{abstract}
Using element-specific X-ray magnetic circular dichroism (XMCD) technique we have studied different magnetic sublattices in a multiferroic Ho$_{0.5}$Nd$_{0.5}$Fe$_{3}$(BO$_{3}$)$_{4}$ single crystal. The XMCD measurements at the \emph{L}$_{2,3}$-edges of Ho and Nd, and at the Fe \emph{K}-edge have been performed at \emph{T}=2~K under a magnetic field up to 17~T applied along the trigonal \emph{c}-axis as well as in the basal \emph{ab}-plane. All three magnetic sublattices are shown to undergo a spin-reorientation transition under magnetic field applied along the \emph{c}-axis. On the contrary, when magnetic field is applied in the \emph{ab}-plane only the holmium atoms exhibit a magnetization jump. Thus, the element-specific magnetization curves revealed the Ho sublattice to be much stronger coupled to the Fe one than the Nd sublattice. The results demonstrate that the Ho$^{3+}$ subsystem plays even more dominant role in magnetic behavior of Ho$_{0.5}$Nd$_{0.5}$Fe$_{3}$(BO$_{3}$)$_{4}$ crystal than in pure HoFe$_{3}$(BO$_{3}$)$_{4}$ crystal.

\begin{description}
\item[PACS numbers]
75.50.Ee, 75.30.Gw, 75.85.+t.
\end{description}
\end{abstract}

\maketitle

\section{\label{sec:level1}INTRODUCTION }

The rare-earth ferroborates \textit{Re}Fe$_{3}$(BO$_{3}$)$_{4}$ (\textit{Re} - rare earth) crystallize in noncentrosymmetrical trigonal \emph{R}32 or \emph{P}3$_{1(2)}$21 space groups, that makes them systems with the coexistence of magnetic order and electric polarization. These materials have been extensively studied since 1970s mainly due to their large magnetoelectric\cite{Zvezdin2006,Liang2011,Zvezdin2005,Kadomtseva2008,Kadomtseva2010,Kadomtseva2012,Mukhin2011,Demidov2011,Kuzmenko2018}, nonlinear optical\cite{Kumar2012,Chen2012} and chiral effects\cite{Usui2014}.

The crystal structure of ferroborates is affected by the rare-earth ion radius leading the structural transition from the \emph{R}32 to the \emph{P}3$_{1}$21 structure for the compounds with a smaller Re ionic radii (e.g. \textit{Re} - Ho, Gd, Tb)\cite{Frolov2016,Hinatsu2003}. The magnetic structure is such that the helicoidal chains formed by edge-shared FeO$_{6}$ octahedra are propagated along the \emph{c}-axis. The magnetic coupling between Fe ions within a single chain is realized through common oxygen atoms with the coupling angle of Fe-O-Fe $\approx$ $100^\circ$. The exchange interaction between nearest magnetic chains takes place via iron-oxygen-rare-earth (Fe-O-\textit{Re}-O-Fe) (\emph{d}$_{Fe-Fe}$ = 3.76 {\AA}) and iron-oxygen-boron Fe-O-B-O-Fe (\emph{d}$_{Fe-Fe}$ = 4.83 {\AA}) pathways. The rare-earth ions are located in a prisms \textit{Re}O$_{6}$, with the inter-ionic distance \emph{d}$_{Re-Re}\sim$ 6 {\AA}. The iron subsystem is antiferromagnetically ordered at \textit{T}$_{N}$ = 30-40 K. The \emph{f}-\emph{f} exchange interaction is weak (\emph{J}$^{Re-Re}$/\emph{k}$_{B}<$1~K) compared with the \emph{J}$_{Fe-Fe}$. The magnetic order in both \textit{Re} and Fe subsystems occurs simultaneously due to the \emph{f}-\emph{d} exchange interaction. Furthermore, depending on the type of \textit{Re} ion, both the easy-plane (EP) and easy-axis (EA) types of magnetic anisotropy can be realized\cite{Chaudhury2009,Pankrats2004,Popov2009}.

\begin{figure}
\includegraphics[width=0.4\textwidth]{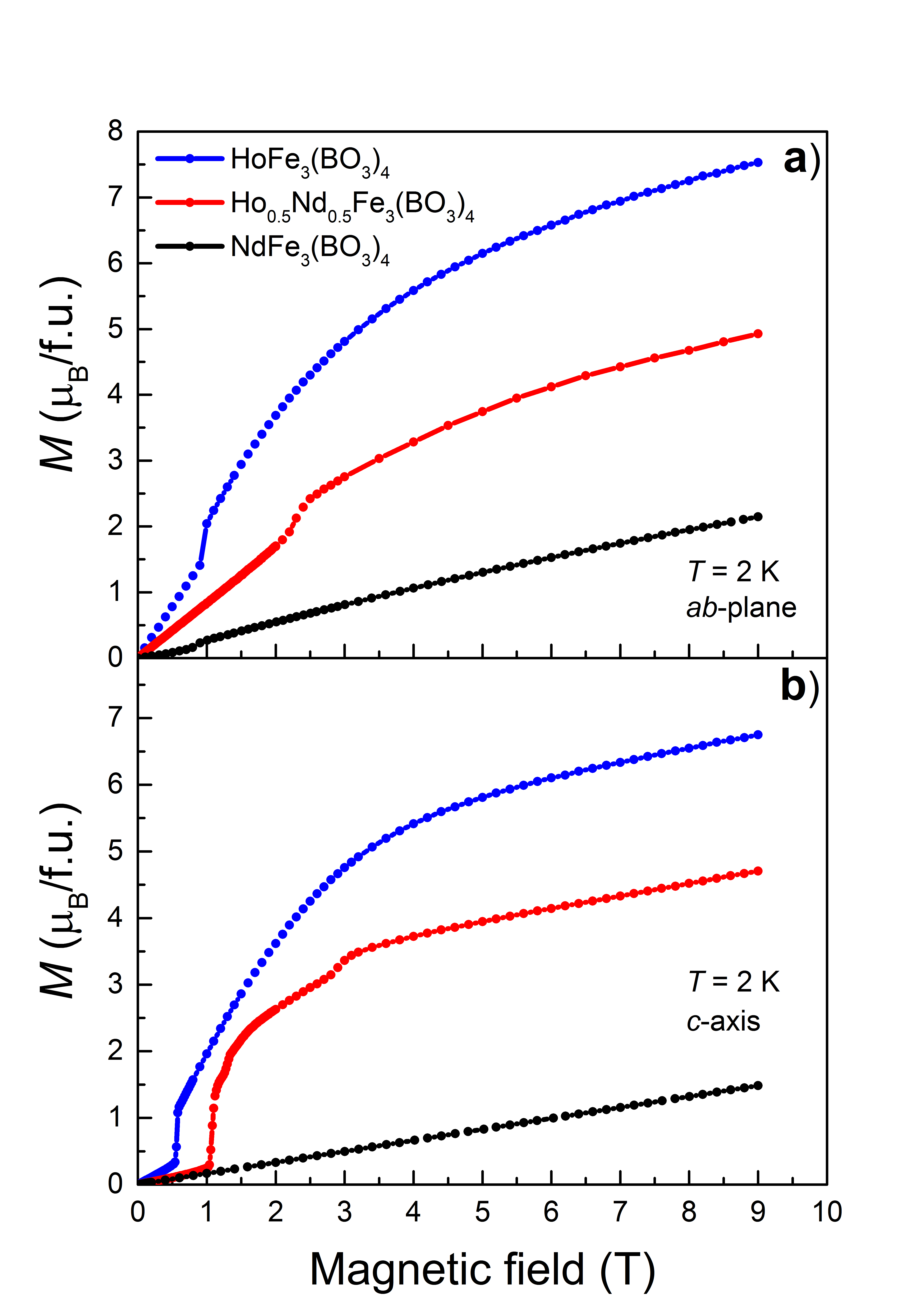}
\caption{(Color online) (a) Field dependences of the magnetizations for Ho$_{1-x}$Nd$_{x}$Fe$_{3}$(BO$_{3}$)$_{4}$ (\emph{x} = 0.0, 0.5, 1.0) for the applied magnetic field in basal \emph{ab}-plane (a) and trigonal \emph{c}-axis (b), \emph{T} = 2 K.}
\label{Fig1}
\end{figure}

In this study, we focused on the Ho$_{0.5}$Nd$_{0.5}$Fe$_{3}$(BO$_{3}$)$_{4}$ single crystal, which is distinguished by its magnetic and magnetoelectric properties from the pure HoFe$_{3}$(BO$_{3}$)$_{4}$ and NdFe$_{3}$(BO$_{3}$)$_{4}$ compounds. The magnetic structures of HoFe$_{3}$(BO$_{3}$)$_{4}$ and NdFe$_{3}$(BO$_{3}$)$_{4}$ were studied using neutron scattering, and resonant/non-resonant magnetic X-ray scattering\cite{Fischer2006,Shukla2012,Partzsch2016,Hamann-Borrero2012,Ritter2008}. In NdFe$_{3}$(BO$_{3}$)$_{4}$ the magnetic anisotropy of Nd$^{3+}$ ions stabilizes the EP magnetic structure with \emph{T}$_{N}$ = 31 K, while in HoFe$_{3}$(BO$_{3}$)$_{4}$ (\emph{T}$_{N}$ = 38 K) the competition between the contributions of holmium and iron sublattices induces the spontaneous spin-reorientation transition from the high-temperature EP to low-temperature EA states at \emph{T}$_{SR}$ = 5 K. The substitution of 50\% Ho$^{3+}$ ions by Nd$^{3+}$ ions leads to the stabilization of the EA phase and the increase in temperature of the spin-reorientation transition up to \emph{T}$_{SR}$ = 9 K \cite{Chaudhury2009}. The field-induced spin-reorientation (spin-flop) transitions observed in the parent compounds are shifted to the high-field region: from $\sim$0.6  T to 1 T for the parallel direction and from $\sim$0.9 T to 2.3 T for the perpendicular direction of magnetic field relative trigonal \emph{c}-axis (Fig.~\ref{Fig1}).

\begin{figure}
\includegraphics[width=0.4\textwidth]{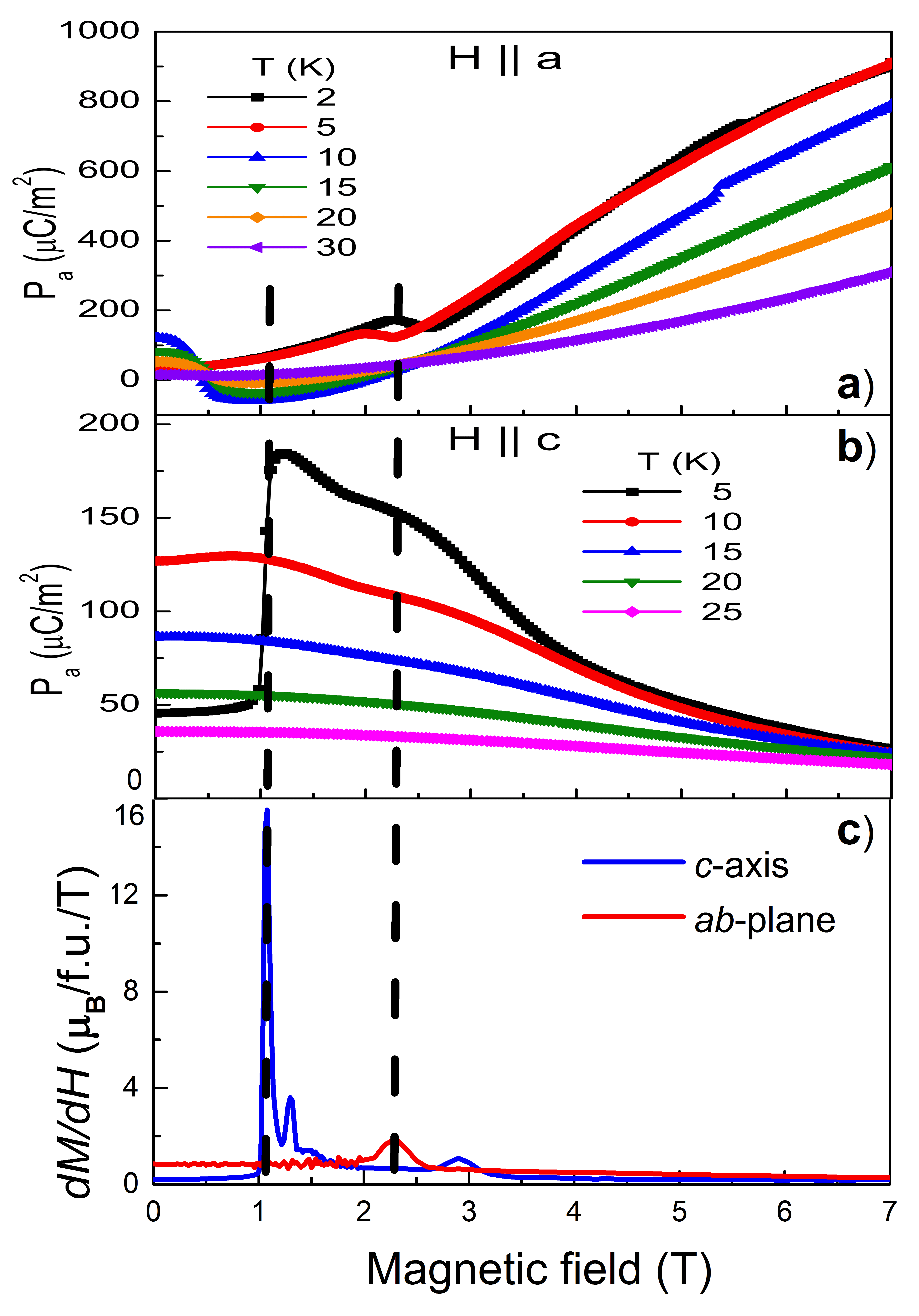}
\caption{(Color online) Magnetic-field dependence of the polarization \emph{P}$_{a}$ at different temperatures. a) \emph{H}$\parallel$\emph{a} axis, b) \emph{H}$\parallel$\emph{c}-axis [data taken from Fig.15 \cite{Chaudhury2009}], c) the derivatives \emph{dM}/\emph{dH} obtained using the magnetic data of Fig.~\ref{Fig1}, which show the onset of spin-flop transitions as a function of applied field (\emph{T} = 2 K). The dashed lines denote the correlation between the spin-flop transitions and the features of the electrical polarization of Ho$_{0.5}$Nd$_{0.5}$Fe$_{3}$(BO$_{3}$)$_{4}$.}
\label{Fig2}
\end{figure}

Moreover, the HoFe$_{3}$(BO$_{3}$)$_{4}$ crystal exhibits a zero-field spontaneous electrical polarization ($\sim$80 $\mu$C/m$^{2}$) below \emph{T}$_{N}$, which is suppressed by the external magnetic fields\cite{Chaudhury2009}. NdFe$_{3}$(BO$_{3}$)$_{4}$ in turn has a much larger (above 300 $\mu$C/m$^{2}$) electric polarization controlled by the magnetic field and the giant quadratic magnetoelectric effect \cite{Zvezdin2006}. Most interestingly, the solid solution of Ho$_{0.5}$Nd$_{0.5}$Fe$_{3}$(BO$_{3}$)$_{4}$ combines the main features observed in the end members: the zero-field spontaneous polarization, which diminishes in the magnetic fields along the \textit{c}-axis similar to HoFe$_{3}$(BO$_{3}$)$_{4}$; and the large value of the polarization (up to 900 $\mu$C/m$^{2}$ at 7 T) induced by the magnetic fields along the \emph{a}-axis exceeding that in NdFe$_{3}$(BO$_{3}$)$_{4}$. The Ho$_{0.5}$Nd$_{0.5}$Fe$_{3}$(BO$_{3}$)$_{4}$ thus demonstrates complex behavior of the temperature and field dependences of the electric polarization\cite{Chaudhury2009}. The sign reversals of the polarization and steplike anomalies correlate with the features at the magnetization curves as shown in the Figure~\ref{Fig2}. The magnetoelectric effect in Ho$_{0.5}$Nd$_{0.5}$Fe$_{3}$(BO$_{3}$)$_{4}$ is therefore likely to may associated with a spin-reorientation transitions in external magnetic fields.

To elucidate a possible relation of this behavior to rare-earth and iron magnetism, a systematic study of the element-specific magnetization has been undertaken. The presence of the three magnetic subsystems in Ho$_{0.5}$Nd$_{0.5}$Fe$_{3}$(BO$_{3}$)$_{4}$ makes this compound an interesting object for the element-specific XMCD study. Our data show that if the magnetic field is applied along the \emph{c}-axis all three (Ho, Nd and Fe) magnetic sublattices undergo the spin-reorientation transition at \emph{H}$^{c}_{sf}$ = 1~T, with the dominant changes in the holmium one. This demonstrates the existence of the magnetic coupling between them. For the applying magnetic field in the \emph{ab}-plane it is only holmium magnetic moment that shows the jump. The Ho$^{3+}$ ions are responsible for the easy-axis magnetic anisotropy inducing spin-flop transitions in the Fe$^{3+}$ and Nd$^{3+}$ sublattices, and participate in the strong magnetic coupling with the Fe$^{3+}$ subsystem. The obtained results show the success in the disentangling of the magnetic contributions of different sublattices in the multiferroic ferroborate.

\section{\label{sec:level2}EXPERIMENTAL TECHNIQUES }

To perform the experimens the Ho$_{1-x}$Nd$_{x}$Fe$_{3}$(BO$_{3}$)$_{4}$ single crystals with a size of 5-7~mm were grown from melt solutions based on bismuth trimolibdate\cite{Chaudhury2009}. The single crystals had a high optical quality, a green color typical for ferroborates, and natural facets. From these crystals, oriented plates were cut, grounded to a thickness of 1-2 mm, and finally polished using diamond powder of 0.25 micron. The crystallographic orientations were identified by a series of X-ray diffraction experiments at the BM01A endstation of the Swiss-Norwegian and ID28 beamlines at the ESRF. The crystal structure of Ho$_{0.5}$Nd$_{0.5}$Fe$_{3}$(BO$_{3}$)$_{4}$ was investigated up to 100 K without detecting any sign of a structural transition. The ratio between enantiomeric twin components for racemic Ho$_{0.5}$Nd$_{0.5}$Fe$_{3}$(BO$_{3}$)$_{4}$ crystal was determined. The refined value of the Flack parameter for Ho$_{0.5}$Nd$_{0.5}$Fe$_{3}$(BO$_{3}$)$_{4}$ at 0.68 ${\AA}$ is 0.27(6). The room temperature lattice parameters are a=9.57280(10) ${\AA}$, b=9.57280(10) ${\AA}$, c=7.58760(10) ${\AA}$, $\alpha$=$\beta$=$90.00^\circ$, $\gamma$=$120.00^\circ$, V=602.161(15) ${\AA}^3$, vary almost linearly with Nd content and are in good agreement with previous published data\cite{Fischer2006, Ritter2008}.

Next the low-temperature macroscopic magnetization measurements in Ho$_{1-x}$Nd$_{x}$Fe$_{3}$(BO$_{3}$)$_{4}$ (\emph{x} = 0.0, 0.5, 1.0) single crystals as a function of applied magnetic fields up to 9 T (Fig.~\ref{Fig1}) were performed using a commercial Physical Property Measurement System (PPMS) from Quantum Design using the Vibrating Sample Magnetometry (VSM) configuration. The magnetization was measured under an external magnetic field applied parallel and perpendicular to the samples' trigonal \emph{c}-axis.

Element-specific XANES and XMCD measurements were then carried out at the Fe \emph{K}- (7112 eV), Ho \emph{L}$_{3,2}$- (8071, 8919 eV), and Nd \emph{L}$_{3,2}$- (6208, 6722 eV) absorption edges at the ESRF ID12 beamline\cite{Rogalev2013}. The APPLE-II undulator and a Si(111) double crystal monochromator were used to collect the spectra at the respective energies. The samples were glued to the sample holder with Apiezon grease. The measurements were done below the spin-reorientation transition at \emph{T} = 2 K. The single crystal was oriented such that the trigonal \emph{c}-axis was parallel and perpendicular to x-ray wave-vector, since x-rays and applied magnetic field are collinear. The XMCD signal was obtained by differences of XANES spectra measured with opposite helicities of the incoming photons at a fixed magnetic field value up to $\pm$17 T. The beam size was approximately 200 $\mu$m, and for each helicity the 24 spectra were recorded and averaged. No radiation damage of the sample was detected. All spectra were measured using the total fluorescence yield detection mode in "backscattering" geometry resulting in the bulk sensitive information. The XMCD spectra at the Fe \emph{K}- and Ho/Nd \emph{L}$_{3,2}$-edges were normalized to the corresponding edge jump of the absorption spectrum. The element-specific magnetization curves were obtained by the measurement of XMCD signal as a function of the field at a fixed energy.

\section{\label{sec:level3}RESULTS }

\begin{figure}
\includegraphics[width=0.4\textwidth]{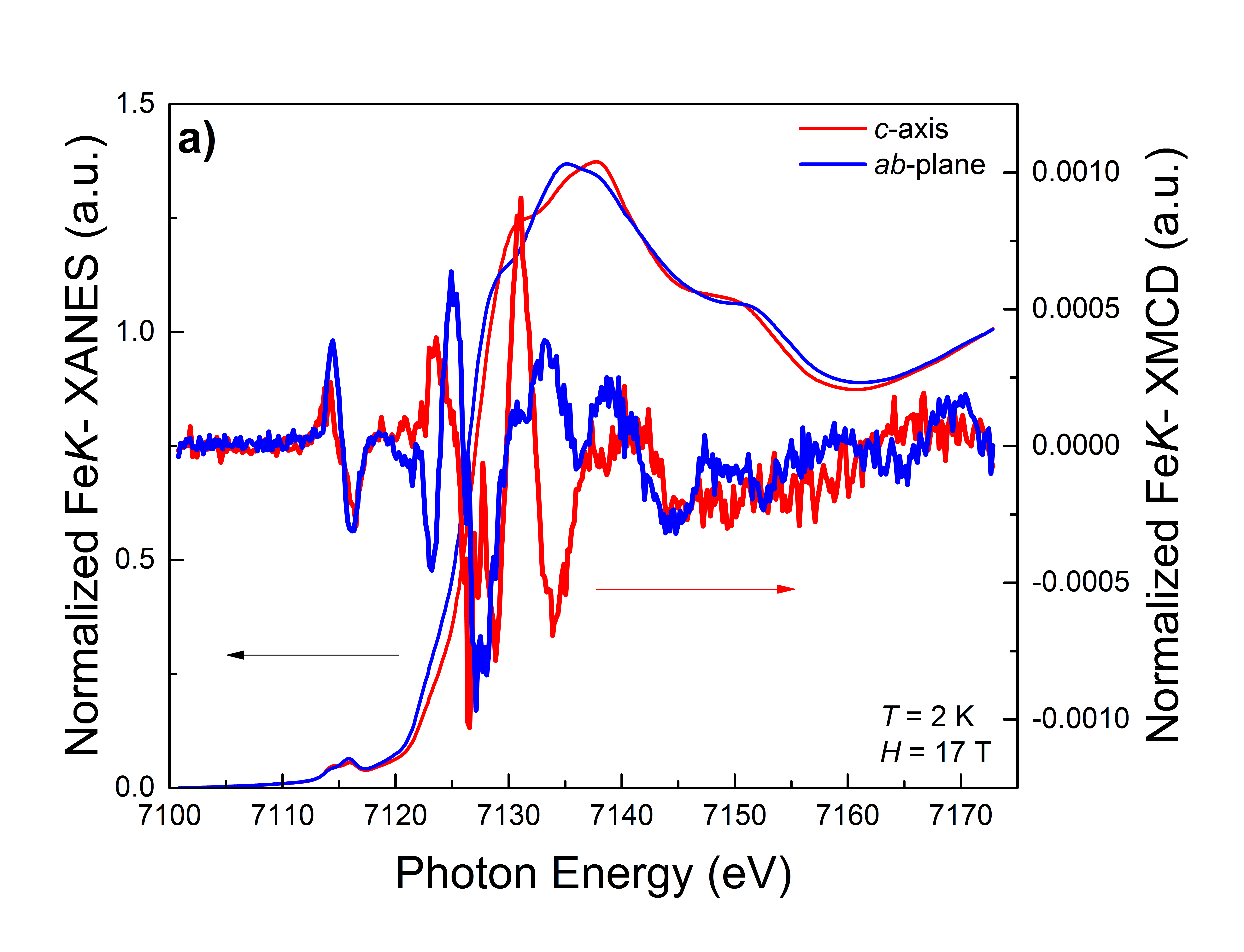}
\includegraphics[width=0.4\textwidth]{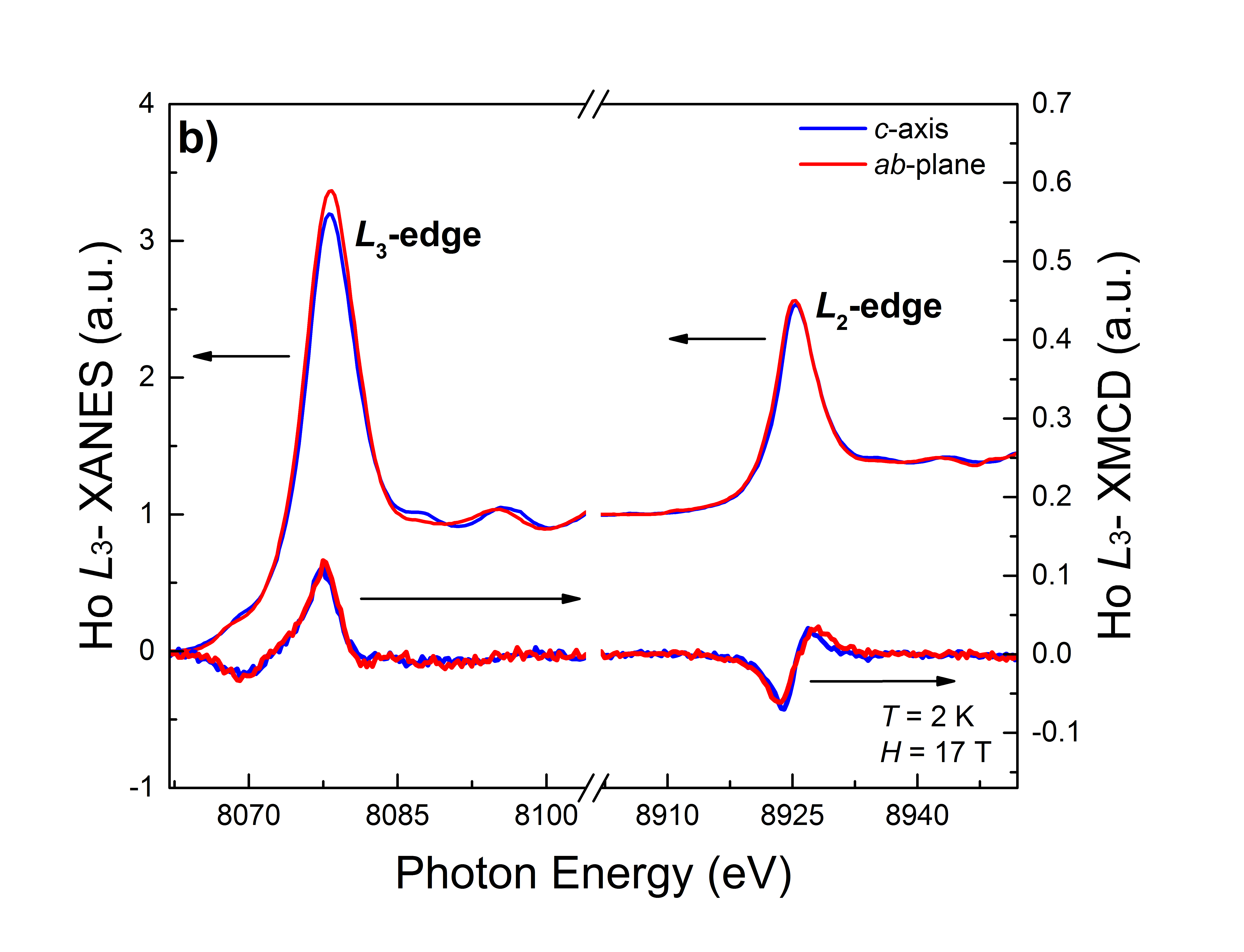}
\includegraphics[width=0.4\textwidth]{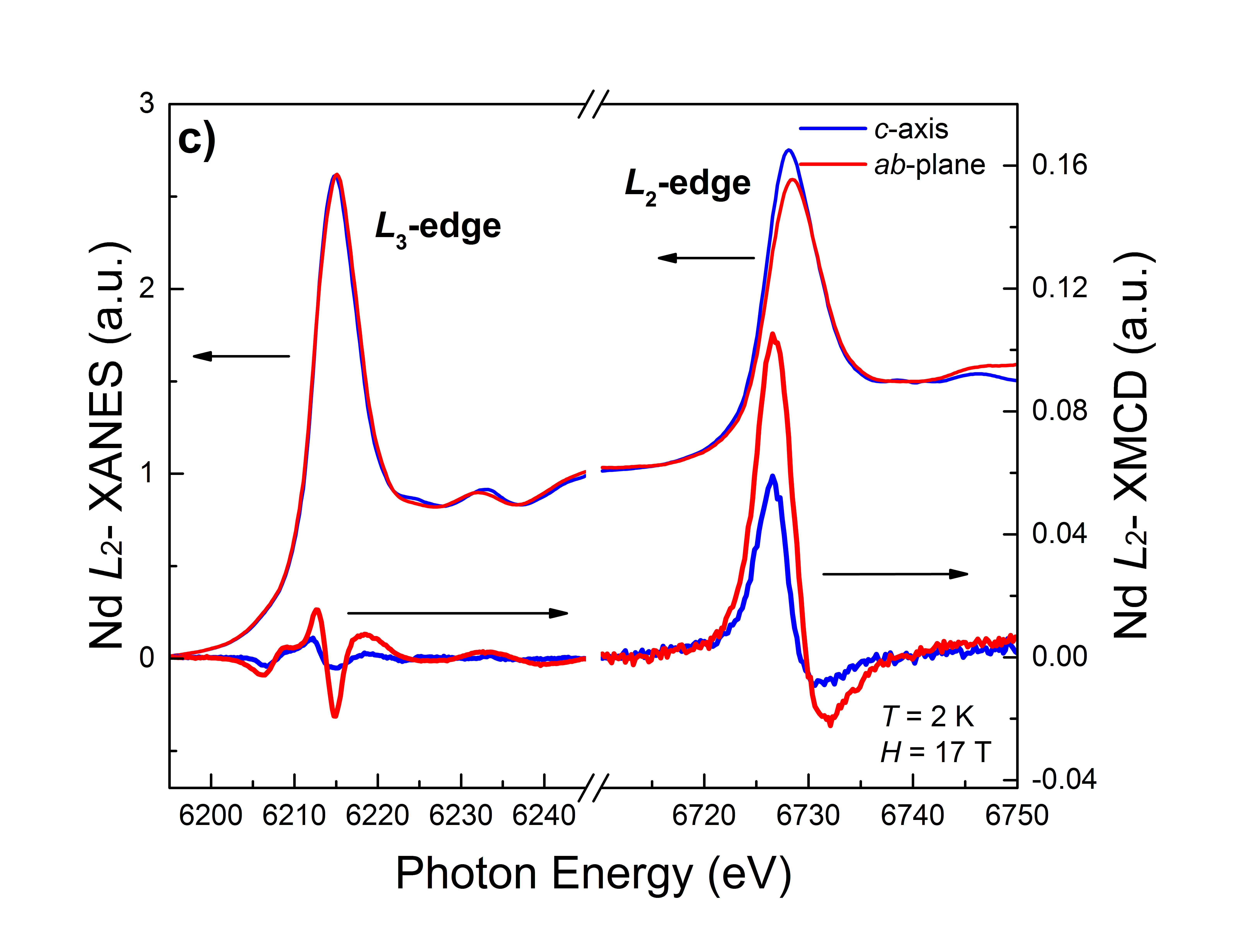}
\caption{(Color online) Normalized XANES/XMCD spectra recorded at 2 K and 17 T for Ho$_{0.5}$Nd$_{0.5}$Fe$_{3}$(BO$_{3}$)$_{4}$ for different orientations of wave vector \emph{k} relative trigonal \emph{c}-axis: a) Fe \emph{K}-edge; b) Ho \emph{L}$_{3,2}$-edges; c) Nd \emph{L}$_{3,2}$-edges.}
\label{Fig3}
\end{figure}

XANES and XMCD spectra at the Fe \emph{K}-edge are shown in Fig.~\ref{Fig3}a. The overall shape of the XMCD signal is interesting. Three spectral energy regions can be identified: a pre-edge region, a rising edge, and a main absorption band. The small double-peak structure between 7110 and 7120 eV is observed in the pre-edge region. The shoulder (7122 eV) in normal incidence is assigned to 1\textit{s}-4\textit{p} of Fe hybridized with 2\textit{p}$_z$ states of the O ligands\cite{DeGroot2008}. Other maxima of the absorption curve in the main absorption band correspond to the transitions to continuum states of \emph{p}-character and are very sensitive to the relative positions of the atoms. A clear dichroic signal can be observed near the edge with the maximum of the order of 8$\cdot$10$^{-4}$ of the edge jump for Fe. The XMCD \emph{K}-edge spectrum is mostly dominated by the excitation of a 1\emph{s} electron to unoccupied 4\emph{p} states. Since the signal is nonzero, it reflects the magnetic polarization of the Fe 4\emph{p} band caused by the intra-atomic exchange interaction with the 3\emph{d} band. So, the Fe \emph{K}-edge XMCD signal is proportional to the magnetization of the 3\emph{d} states. The probing depth at the \emph{K}-edge is considerably larger than the escape depth of electrons that are usually used for detection of X-ray absorption and XMCD spectra at the \emph{L}$_{3,2}$-edges of 3\emph{d} transition metals. Accordingly, the present data are representative of the bulk, that is of importance to a study of magnetic dielectric oxides.

XANES and XMCD spectra collected at the Ho and Nd \emph{L}$_{3,2}$-edges are shown in Fig.~\ref{Fig3}b and ~\ref{Fig3}c respectively. The Ho and Nd XANES spectra were normalized to one at the \emph{L}$_{3}$- and half at the \emph{L}$_{2}$-edge to reflect the 2:1 ratio of the initial state at these edges (2\emph{p}$_{3/2}$ and 2\emph{p}$_{1/2}$, respectively). Each of these spectra can be decomposed into three main features: the first corresponds to 2\emph{p}$\rightarrow$5\emph{d} electronic transitions; second, a steplike edge feature that is associated with 2\textit{p} $\rightarrow$ continuum electronic excitations; and, third, a series of smaller "fine structure" oscillations that arise from the backscattering of photoelectrons by neighboring atoms. Dipolar selection rules make the magnetic signal at the rare-earth \emph{L}$_{3,2}$-absorption edges sensitive to the spin polarization of the intermediate 5\emph{d} level as a result of a 4\emph{f}-5\emph{d} exchange interaction\cite{Galera2008}. 

The XMCD signal obtained across the two edges shows different intensities with a strong dichroic magnetic signal. The \emph{L}$_{3}$-edges XMCD spectra of both Ho and Nd consist of a negative dip followed by a main positive peak above the Fermi energy. Such a negative dip is very small for the Nd case. This spectral feature has been associated to a quadrupolar transition (2\emph{p}$\rightarrow$4\emph{f}) that should be present at the \emph{L}$_{3}$-edge spectra of heavy rare-earth and its magnitude is small for light lanthanide, i.e. Nd\cite{Bartolome1997,Bartolome1999}. The Ho \emph{L}$_{2}$-edge XMCD spectra consists of a main negative peak above the edge and a smaller positive peak at higher energy. The amplitude of the \emph{L}$_{2}$-edge XMCD spectra decreases from Ho to Nd as well. We can assume that the Ho moments in the Ho$_{0.5}$Nd$_{0.5}$Fe$_{3}$(BO$_{3}$)$_{4}$ sample are polarized in the direction of applied field, as follows from the positive \emph{L}$_{3}$- and negative \emph{L}$_{2}$-peaks\cite{Bartolome1997,Bartolome1999}.

\begin{figure}
\includegraphics[width=0.4\textwidth]{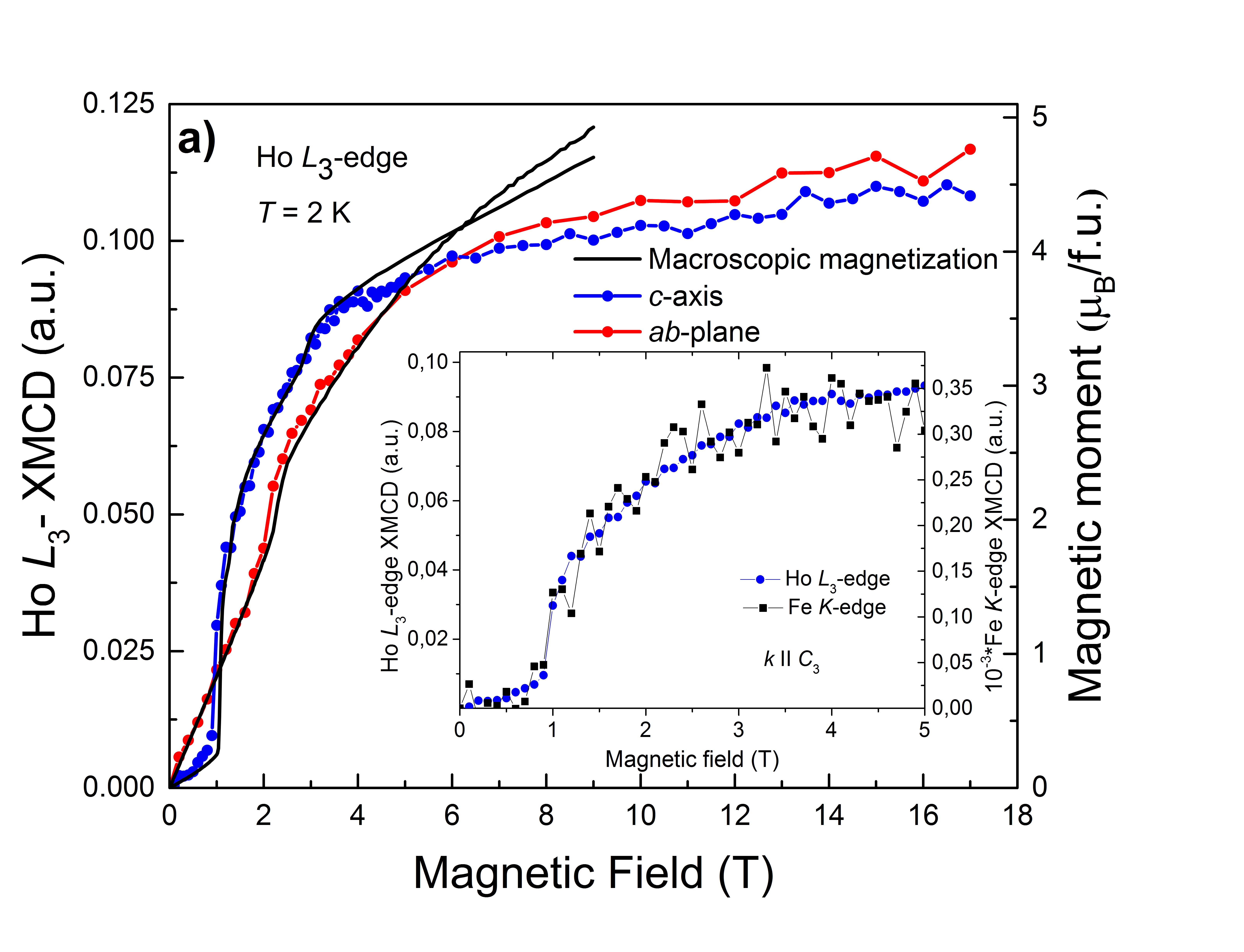}
\includegraphics[width=0.4\textwidth]{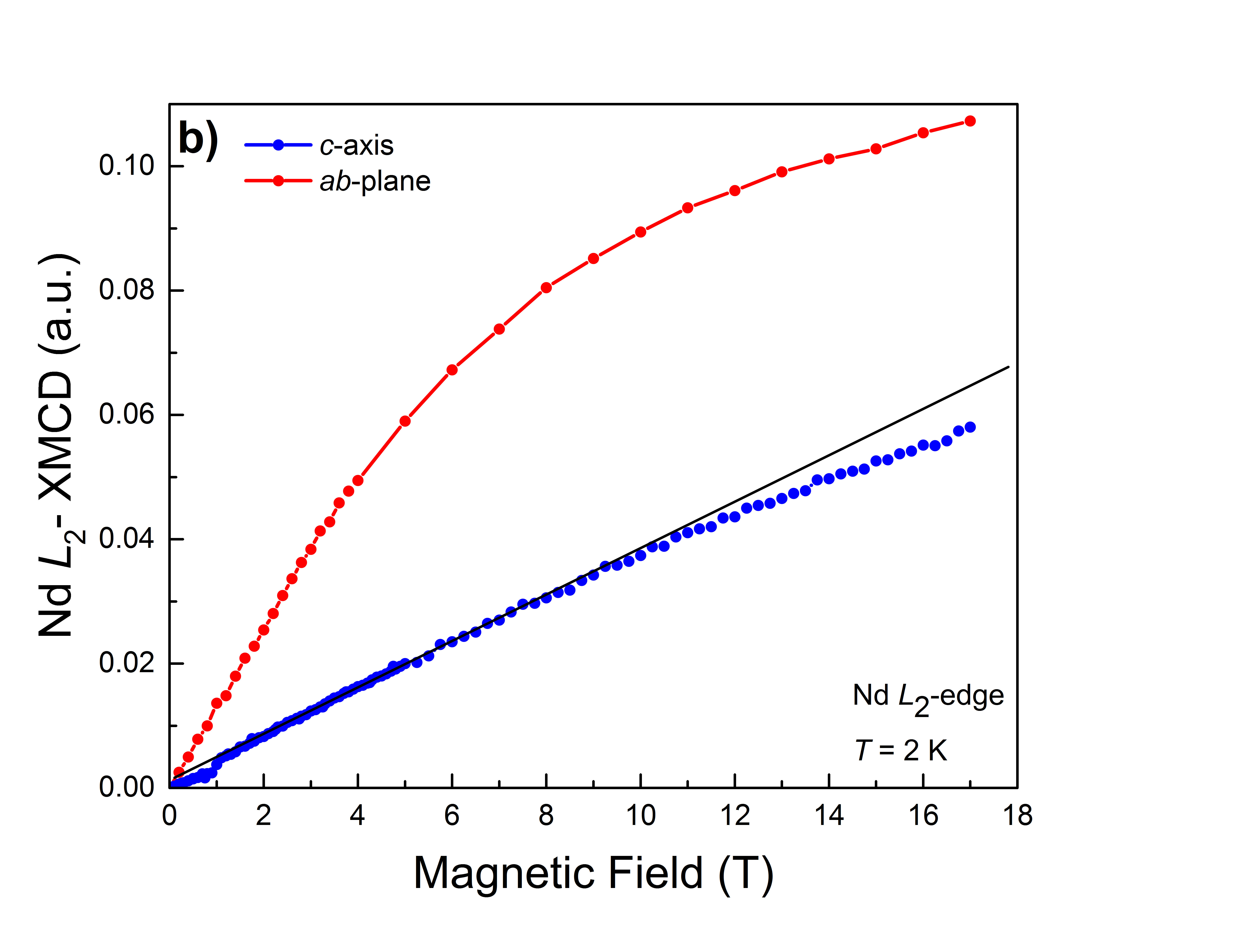}
\caption{(Color online) (a) The element-specific magnetization curves of Ho$_{0.5}$Nd$_{0.5}$Fe$_{3}$(BO$_{3}$)$_{4}$ as a function of applied field: a) Ho \emph{L}$_{3}$-edge and bulk magnetization curves. The inset: the comparison of Ho \emph{L}$_{3}$- and Fe \emph{K}-edges XMCD magnetization curves for the applied field along c-axis; b) Nd \emph{L}$_{2}$-edge. The inset shows the linear dependence of the Nd \emph{L}$_{2}$-edge XMCD magnetization on the applied field along \emph{c}-axis.}
\label{Fig4}
\end{figure}

It is important to note that the evolution with the field of the XMCD signal at the \emph{L}$_{3,2}$-edges can be considered directly proportional to the local magnetization from the 3\emph{d} or 4\emph{f} states. The specific energies that correspond to the maximum amplitude of the XMCD signals at the \emph{L}$_{3}$-edge of Ho (\emph{E}$^{\emph{L}_3}_{Ho}$ = 8077.13 eV), and the \emph{K}-edge of Fe (E$^K_{Fe}$=7124.83 eV) were selected. Due to the small value of the XMCD signal at \emph{L}$_{3}$-edge of Nd, the \emph{L}$_{2}$-edge was selected for XMCD measurement (E$^{L_2}_{Nd}$ = 6726.16 eV).
The element-specific magnetization curves of Ho$_{0.5}$Nd$_{0.5}$Fe$_{3}$(BO$_{3}$)$_{4}$ at the Fe \emph{K}-, as well as at the Ho, Nd \emph{L}$_{3,2}$-edges in comparison with macroscopic magnetization are shown in Fig.~\ref{Fig4}. Depending on their direction, the external magnetic fields have a drastic effect on the magnetic behavior of rare-earth and iron sublattices.

\emph{Ho L$_{3}$-edge.} The field dependence of the Ho \emph{L}$_{3}$-edge magnetization reveals a sharp increase of the magnetic moment at the critical fields of \emph{H}$^{\emph{c}}_{sf}$=1 T for applied magnetic field along the trigonal \emph{c}-axis. The magnetization jump reduces in value, becomes less sharp and is shifted toward higher fields (\emph{H}$^{\emph{ab}}_{sf}$=2.2 T) for magnetic field applied in the basal \emph{ab}-plane. These jumps are related to the holmium magnetic moments reorientations. In the spin-flop phase (\emph{H}$<$\emph{H}$_{sf}$) there is a linear contribution to holmium magnetization with the initial susceptibility in the basal \emph{ab}-plane more large than that along \emph{c}-axis. It is worth noting that the Ho \emph{L}$_{3}$-edge magnetization obtained from the XMCD data matches well the bulk magnetization measurements at the fields \emph{H}$<$5 T and tends to the saturation at larger fields. At the fields \emph{H}$\sim$6 T the intersection of the XMCD magnetization curves measured along and perpendicular to the \emph{c}-axis takes place.

\emph{Nd L$_{2}$-edge.} The Nd \emph{L}$_{2}$-edge magnetization shows a strong anisotropy (Fig.~\ref{Fig4}b). The applying magnetic field along trigonal \emph{c}-axis causes a visible jump in the magnetization at \emph{H}$^{\emph{c}}_{sf}$=1 T. There is a linear field dependence of the magnetization for the fields \emph{H}$^{\emph{c}}_{sf}$$<$\emph{H}$<$9 T and the tendency to the flattening at \emph{H}$>$9 T. In contrast, for applied magnetic fields in the basal \emph{ab}-plane the nonlinear behavior of the magnetization without any features is observed up to highest fields. At magnetic field of 17 T the Nd \emph{L}$_{2}$-edge XMCD signal in \emph{ab}-plane is twice of that along \emph{c}-axis.

\emph{Fe K-edge.} The Fe \emph{K}-edge magnetization as a function of the applied field is presented in the inset to Fig.~\ref{Fig4}a. The Fe K-edge XMCD signal is two orders value smaller than that at Ho \emph{L}$_{3}$-edge, confirming the antiferromagnetic ordering of Fe$^{3+}$ magnetic moments as presumed from macroscopic magnetization study. The magnetic field applied along the \emph{c}-axis induces a sharp jump in the iron magnetization at \emph{H}$^{\emph{c}}_{sf}$ = 1~T similar that observed for the rare-earth magnetization curves. At the fields above \emph{H}$^\emph{c}_{sf}$ the iron magnetization follows to the holmium magnetization and tends to the saturation indicating that these atoms are strongly magnetically coupled. The magnitude of the Fe \emph{K}-edge XMCD signal in \emph{ab}-plane is commensurable with that of the noise. For this reason these data does not discussed.

\section{\label{sec:level4}DISCUSSION }

For both orientations of the magnetic fields the Ho \emph{L}$_{3}$-, Nd \emph{L}$_{2}$-, and Fe \emph{K}-edge magnetization curves tend to zero as the magnetic field tends to zero. This observation agrees with the assumption of the polarization of rare-earth sublattices by antiferromagnetic interactions with the iron subsystem (\textit{Re}$^{3+}$-O-Fe$^{3+}$). The each of the rare-earth sublattice (Ho and Nd) should be divided into two magnetic sublattices with magnetic moments oriented oppositely. At \emph{T}$<$\emph{T}$_{SR}$ the magnetic moments of all three magnetic sublattices are presumably oriented along trigonal \emph{c}-axis. For this reason the Fe \emph{K}-edge magnetization shows very small longitudinal susceptibility prior to the spin-flop transition. As applied field along \emph{c}-axis increases the contribution of the rare-earth sublattice, the magnetic moment opposed the field is decreased determining the character of the magnetization in the spin-flop phase. At \emph{H}$^{\emph{c}}_{sf}$=1 T the magnetic moments reorientation occurs in a jump. The observation of the XMCD magnetizations jumps indicates that a magnetic field applied along \emph{c}-axis changes the orientations of the magnetic moments of all three magnetic sublattices and primarily concerned the holmium one. If the external field is applied in the basal \emph{ab}-plane only the Ho \emph{L}$_{3}$-edge magnetization curve shows the jump at \emph{H}$^{\emph{ab}}_{sf}$=2.2 T.

In the flop phase (\emph{H}$>$\emph{H}$_{sf}$) the holmium and neodymium sublattices magnetizations measured via XMCD have a quite different behaviors. The Ho \emph{L}$_{3}$-edge magnetization tends to the saturation, demonstrating the small cant angle of the magnetization curve (perpendicular susceptibility). This lack of saturation is probably due to the weak \emph{f}-\emph{d} exchange interaction. For the same reason, the magnetic moments of the holmium sublattices undergo the spin-reorientation transition more rapidly than those of the iron sublattices. At the magnetic fields above spin-flop transition the Fe magnetic moments gain a small cant toward the external field direction under the action of the effective magnetic field. It was previously proposed that the \emph{f}-\emph{d} exchange interaction does not affect the process of rotation the Fe magnetic moments at the magnetic fields \emph{H}$>$\emph{H}$_{sf}$, giving rise to linear dependence on the field. However, XMCD measurements in Ho$_{0.5}$Nd$_{0.5}$Fe$_{3}$(BO$_{3}$)$_{4}$ show strong magnetic coupling between the holmium and iron sublattices, which is reflected in the lack of the saturation in the former and the nonlinear behavior of the magnetization in the latter.

The Nd magnetization curves show strong anisotropy, which is manifested in the difference in the magnitudes of XMCD signals and in the form of the magnetization curves obtained for two magnetic field orientations. For the applied field in the \emph{ab}-plane the magnitude of the XMCD signal reaches ~0.11 at 17 T and is comparable with that found in permanent magnet Nd$_{2}$Fe$_{14}$B\cite{Menushenkov2017}. At the same time the Nd XMCD dichroic signal along trigonal \emph{c}-axis reaches the smaller magnitude of $\sim$0.6. This observed difference reflects the easy-plane anisotropy of Nd$^{3+}$ ion.

\begin{figure}
\includegraphics[width=0.4\textwidth]{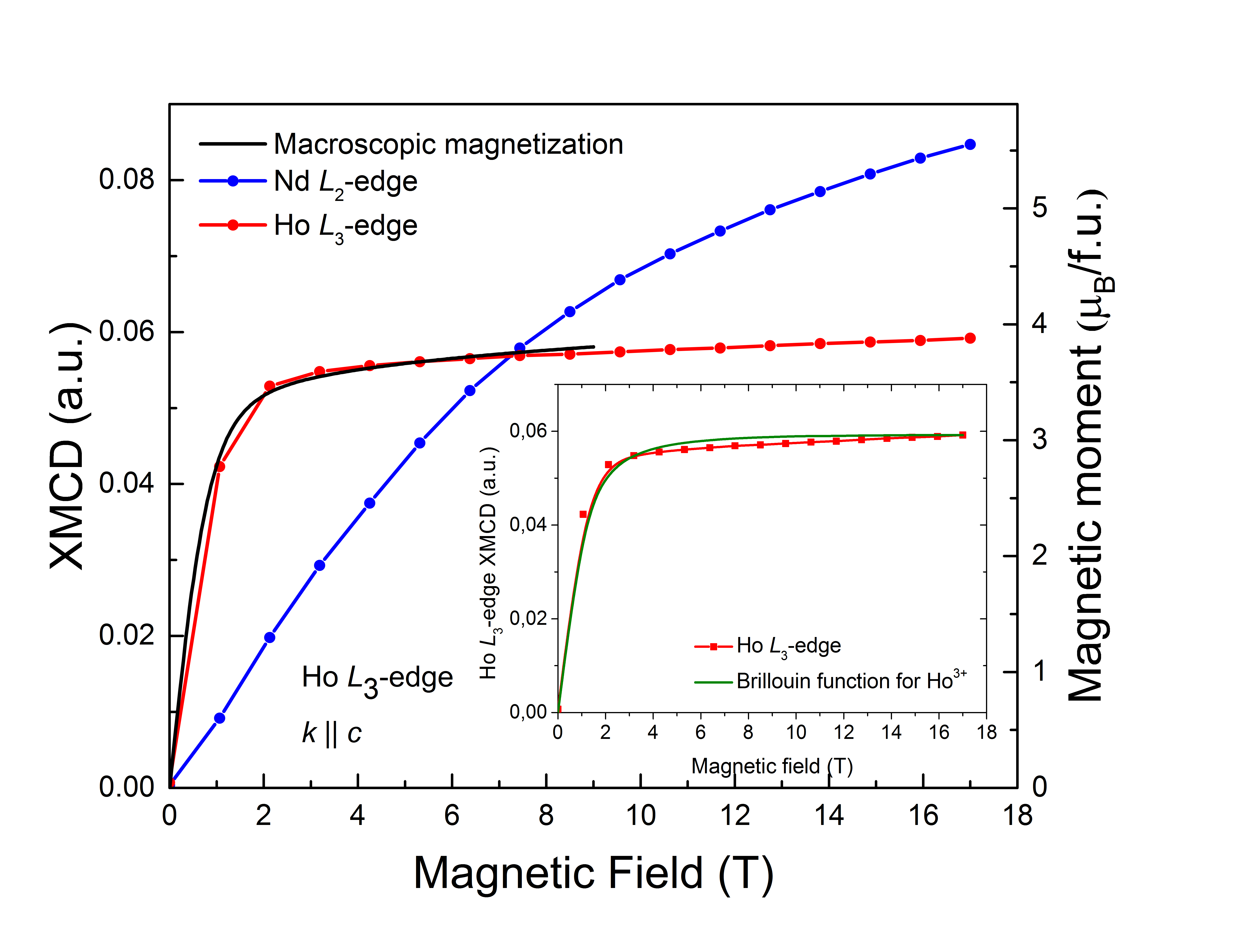}
\caption{(Color online) The macroscopic and element-specific magnetization curves of Ho$_{0.5}$Nd$_{0.5}$Al$_{3}$(BO$_{3}$)$_{4}$  at Ho \emph{L}$_{3}$- and Nd \emph{L}$_{2}$-edges as a function of the applied field along \emph{c}-axis, \emph{T}=2 K The inset shows the fit using the Brillouin function.}
\label{Fig5}
\end{figure}

The effect of the \emph{f}-\emph{d} exchange interaction on the rare-earth sublattices can be studied by measuring the element-specific magnetization curve at Ho \emph{L}$_{3}$- and Nd \emph{L}$_{2}$-edges in Ho$_{0.5}$Nd$_{0.5}$Al$_{3}$(BO$_{3}$)$_{4}$. This compound is paramagnet showing the large magnetoelectric effect $\Delta$\emph{P}$\approx$1400 $\mu$C/m$^{2}$ at 9 T exceeding known values in the ferroborates. The element-specific magnetization curves in Ho$_{0.5}$Nd$_{0.5}$Al$_{3}$(BO$_{3}$)$_{4}$ measured at the magnetic field applied along trigonal \emph{c}-axis and at \emph{T}=2 K are shown in Fig.~\ref{Fig5}. When the Fe ion is replaced by Al ion the amplitude of the XMCD signal at Ho \emph{L}$_{3}$-edge decreases. Indeed, in the absence of the exchange interaction with the iron subsystem the Ho magnetization curve clearly shows the saturation at fields above 6 T while the Nd magnetization curve monotonically increases up to the highest field. Considering that the XMCD at the rare-earth \emph{L}$_{3,2}$-absorption edges is exclusively owing to the rare-earth and is proportional to the 4\emph{f} electrons we can assume that in the case of Ho$_{0.5}$Nd$_{0.5}$Al$_{3}$(BO$_{3}$)$_{4}$ the Ho and Nd XMCD contributions are close to the free-ion ones. We fitted the Ho \emph{L}$_{3}$-edge magnetization using the Brillouin function assuming \emph{J}=8 and \emph{g}=1.25 close to free Ho$^{3+}$ ion. The agreement with the XMCD data is quite good with a saturation of $\sim$0.06. However, using the free-ion parameters for the Brillouin function of Nd \emph{L}$_{2}$-edge magnetization leads to the disagreement. The main difference is that free Ho$^{3+}$ ion with \emph{f}$^{10}$ configuration has the ground state $^{5}$\emph{I}$_8$ (\emph{S}=2, \emph{L}=6) separated by $\Delta$=5000 cm$^{-1}$ from the first excited state, while the ground state of Nd$^{3+}$ ion with \emph{f}$^{3}$ configuration is $^{4}$\emph{I}$_{9/2}$ (\emph{S}=3/2, \emph{L}=6) and is separated by smaller gap of $\Delta$$\approx$1800 cm$^{-1}$ from the first excited state $^{4}$\emph{I}$_{11/2}$ leading to the admixing to the ground state. The next important finding is the Ho magnetization is well agreed with bulk magnetization indicating the dominant contribution to the magnetism and magnetocrystalline anisotropy of Ho$_{0.5}$Nd$_{0.5}$Al$_{3}$(BO$_{3}$)$_{4}$ similar to Ho$_{0.5}$Nd$_{0.5}$Fe$_{3}$(BO$_{3}$)$_{4}$.

So, one can conclude that the Ho$^{3+}$ magnetic sublattice is responsible for the spin-reorientation transitions in the Ho$_{0.5}$Nd$_{0.5}$Fe$_{3}$(BO$_{3}$)$_{4}$ antiferromagnet and plays a crucial role in the magnetism of the system. The Ho$^{3+}$ magnetic moment determines the direction of the Fe$^{3+}$ and consequently of the Nd$^{3+}$ magnetic moments, causing them to be guided along the trigonal \emph{c}-axis at low temperatures and involves them to the spin-orientation process. The intersection of the bulk magnetization curves of Ho$_{0.5}$Nd$_{0.5}$Fe$_{3}$(BO$_{3}$)$_{4}$ at $\sim$6 T is similar to that observed at the HoAl$_{3}$(BO$_{3}$)$_{4}$ and Ho$_{0.5}$Nd$_{0.5}$Al$_{3}$(BO$_{3}$)$_{4}$ at $\sim$7.5 T and was not observed in HoFe$_{3}$(BO$_{3}$)$_{4}$. This feature is due to the Ho$^{3+}$ magnetic contribution as shown from the XMCD data in Fig.~\ref{Fig4}a and indicates the qualitative changes of this contribution comparing with HoFe$_{3}$(BO$_{3}$)$_{4}$ due to the crystal field effect.

\begin{figure}
\includegraphics[width=0.4\textwidth]{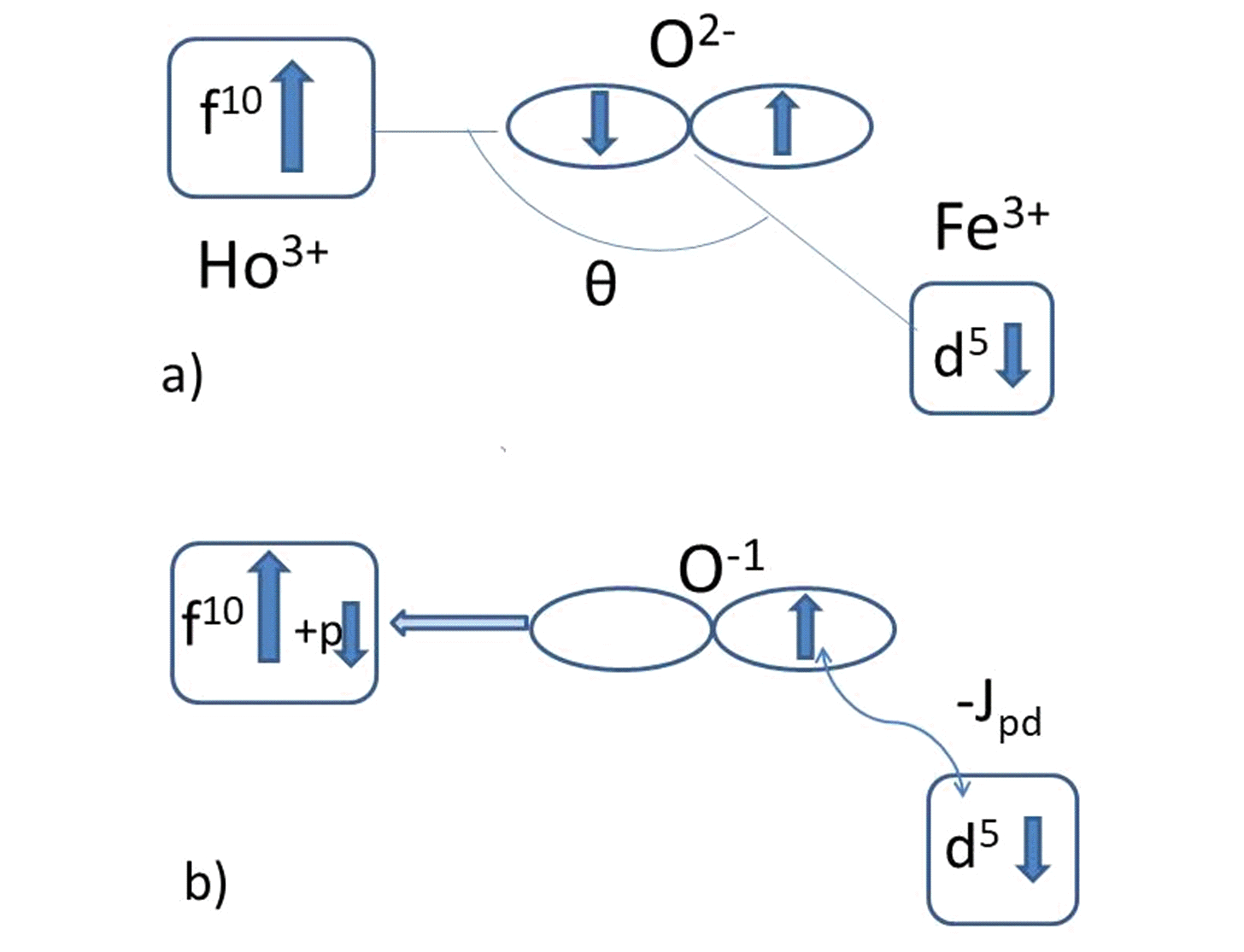}
\caption{(Color online) Scheme of the superexchange interaction for Ho$^{3+}$-O$^{2-}$-Fe$^{3+}$ chain with the angle $\theta$. a) The initial state. b) the oxygen spin down electron transfers to Ho ion while spin up electron has exchange coupling -\emph{J}$_{\emph{pd}}$ with Fe ion spin.}
\label{Fig6}
\end{figure}
\begin{figure}
\includegraphics[width=0.4\textwidth]{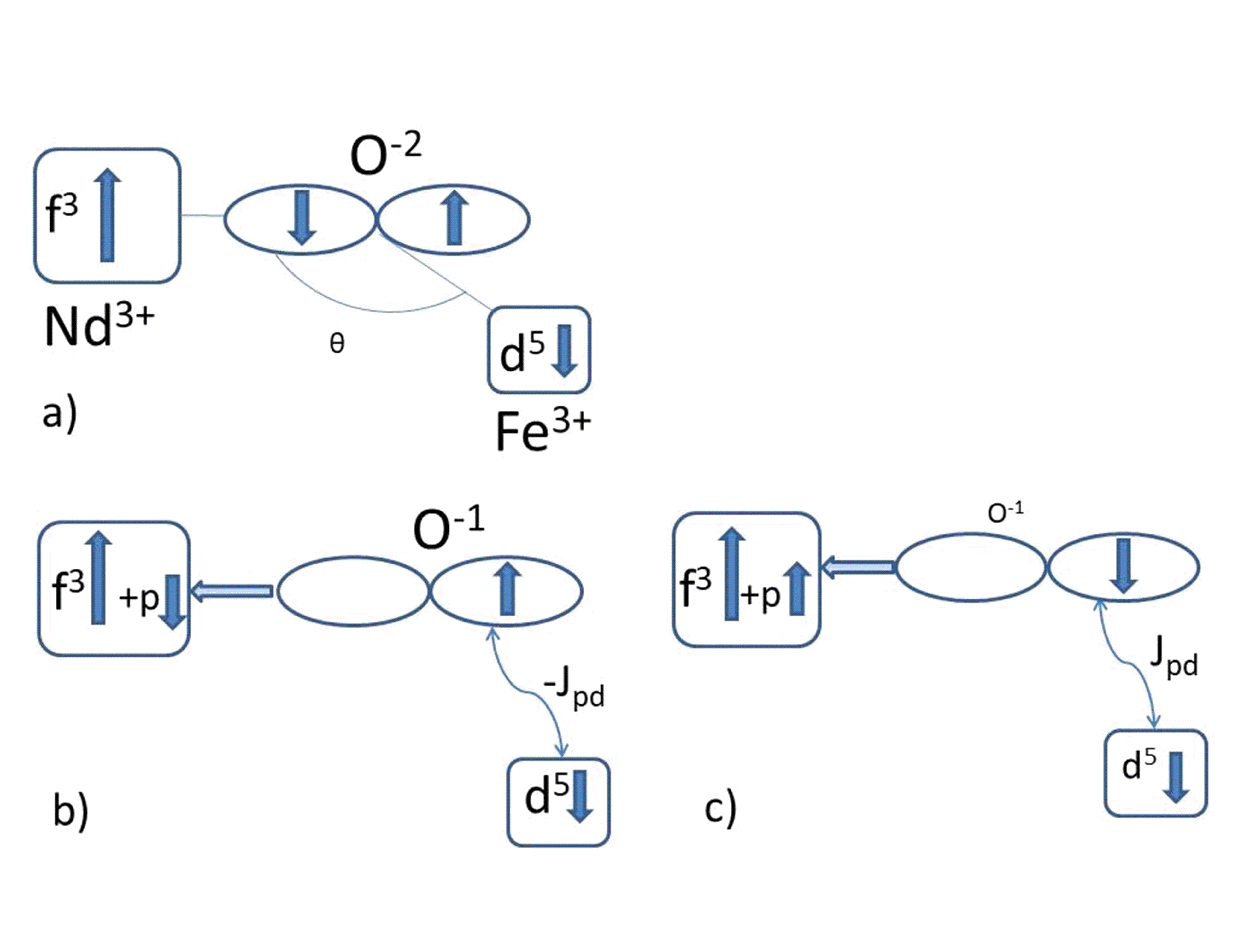}
\caption{(Color online) Scheme of the superexchange interaction for Nd$^{3+}$-O$^{2-}$-Fe$^{3+}$ chain with the angle $\theta$. a) The initial state. b) The oxygen spin down electron transfers to Nd ion while spin up electron has exchange coupling -\emph{J}$_{\emph{pd}}$ with Fe ion spin. c) The oxygen spin up electron transfers to Nd ion while spin down electron has exchange coupling \emph{J}$_{\emph{pd}}$ with Fe ion spin.}
\label{Fig7}
\end{figure}

The neodymium sublattice coupled by comparatively weak exchange interaction with the iron sublattice undergoes considerably smaller change in the magnetic moment. The effect of the Nd$^{3+}$ magnetic subsystem is mainly manifested in the linear contribution to the macroscopic magnetization along \emph{c}-axis and in the extra contribution to the magnetization in the \emph{ab}-plane. To gain insight into the reduction of the neodymium contribution in the Ho$_{0.5}$Nd$_{0.5}$Fe$_{3}$(BO$_{3}$)$_{4}$ solid solution estimations of the \emph{f}-\emph{d} superexchange interactions are needed. The qualitative analysis of the superexchange interaction for Ho$^{3+}$-O$^{2-}$-Fe$^{3+}$ and Nd$^{3+}$-O$^{2-}$-Fe$^{3+}$ chains is given below. Following to the Kramers-Anderson model, the superexchange interaction for Ho$^{3+}$-O$^{2-}$-Fe$^{3+}$ chain with an angle $\theta$$\sim$$120^\circ$ is forming as illustrated in Fig.~\ref{Fig6}. According to the Pauli principle only spin down oxygen electron can transfer to Ho$^{3+}$ ion with more than half-filled \emph{f}-shell. The oxygen spin up electron has exchange coupling -\emph{J}$_{\emph{pd}}$ with Fe ion spin. For less than half-filled \emph{f}-shell of Nd$^{3+}$ ion both oxygen electrons with spin up and spin down can transfer to Nd$^{3+}$ ion (Fig.~\ref{Fig7}). Spin down transferred electron result in \emph{-J}$_{\emph{pd}}$ coupling (Fig.~\ref{Fig7}b), while the oxygen spin up electron transfer result in opposite sign \emph{-J}$_{\emph{pd}}$ coupling (Fig.~\ref{Fig7}c). Thus both spin channels for Nd$^{3+}$-O$^{2-}$-Fe$^{3+}$ chain partially compensate each other and resulting superexchange interaction for Nd$^{3+}$-O$^{2-}$-Fe$^{3+}$ chain is weaker that for Ho$^{3+}$-O$^{2-}$-Fe$^{3+}$ one.

\section{\label{sec:level5}CONCLUSIONS }

The Fe \emph{K}-edge, Ho \emph{L}$_{3,2}$-edge and Nd \emph{L}$_{3,2}$-edge XMCD spectra have been measured at \emph{T}=2 K under magnetic field $\pm$17 T for the Ho$_{0.5}$Nd$_{0.5}$Fe$_{3}$(BO$_{3}$)$_{4}$ single crystal. The external magnetic field was applied along the trigonal \emph{c}-axis and in the basal \emph{ab}-plane. It was found, that all three magnetic sublattices undergo spin-reorientation transition at \emph{H}$^{\emph{c}}_{sf}$=1~T if the magnetic field is applied along \emph{c}-axis, and only holmium subsystem shows the sharp increase of the magnetic moment at \emph{H}$^{\emph{ab}}_{sf}$=2.2 K for the applied magnetic field in the \emph{ab}-plane. The main conclusion of this study is to the Ho and Fe magnetic sublattices are strongly coupled and the magnetic properties including the magnetocrystalline anisotropy in the Ho$_{0.5}$Nd$_{0.5}$Fe$_{3}$(BO$_{3}$)$_{4}$ single crystal are due to anisotropy of the Ho sublattice. This conclusion seems unexpected taking into account the significant substitution (50\%) of holmium ions by neodymium ones, which have different anisotropy signs. Nevertheless the assumption that the Ho$^{3+}$ ion magnetic anisotropy is enhanced in Ho$_{0.5}$Nd$_{0.5}$Fe$_{3}$(BO$_{3}$)$_{4}$ in compare with HoFe$_{3}$(BO$_{3}$)$_{4}$ can explain the observed stabilization an easy-axis magnetic structure at higher temperature (\emph{T}$_{SR}$) and magnetic field (\emph{H}$_{sf}$). The spin-flop transitions observed at the macroscopic magnetization curve and associated anomalies of electrical polarization are shown to be due to the Ho$^{3+}$ magnetic moment reorientation from the easy-axis to easy-plane state. The reasons of the electric polarization changes found in Ho$_{0.5}$Nd$_{0.5}$Fe$_{3}$(BO$_{3}$)$_{4}$  system relative the parent samples HoFe$_{3}$(BO$_{3}$)$_{4}$ and NdFe$_{3}$(BO$_{3}$)$_{4}$ is probably due to the increase in the Ho sublattice contribution as well as to the qualitative changes of this contribution due to the crystal field effect.

\begin{acknowledgments}
The study was supported in part by the grants of the Russian Foundation for Basic Research (project nos. 16-32-60049, 17-02-00826) and the Foundation for Assistance to Small Innovative Enterprises (FASIE, UMNIK program). The authors thank the BM01A/ID28/ID12 beamlines staffs for help during the experiment and also gratefully acknowledge the beamtime provision (proposal HC-1804) by the ESRF.
\end{acknowledgments}


\end{document}